\documentclass[journal]{IEEEtran}
\usepackage{graphicx,amssymb,lineno}
\usepackage{amsmath,amsfonts,amssymb}

\usepackage{algorithmicx}
\usepackage[usenames]{color}
\usepackage{float}
\usepackage[linesnumbered,ruled,vlined]{algorithm2e}
\usepackage{graphicx,graphics,color,epsfig,subfigure,graphpap,rotate}
\usepackage{times, verbatim, subfigure, epsfig, graphicx, latexsym, amsmath}
\usepackage{url}
\usepackage{subfigure}
\usepackage{cite}
\begin{document}
\title{Mobility-aware Caching Scheduling for Fog Computing in mmWave Band}

\author{Yong~Niu,~\IEEEmembership{Member,~IEEE,}
        Yu Liu,
        Yong~Li,~\IEEEmembership{Senior Member,~IEEE,}
        Zhangdui Zhong,~\IEEEmembership{Senior Member,~IEEE,}
        Bo~Ai,~\IEEEmembership{Senior Member,~IEEE},
        and Pan Hui,~\IEEEmembership{Fellow,~IEEE}
\thanks{Y. Niu, Z. Zhong, and B. Ai are with the State Key Laboratory of Rail Traffic
Control and Safety, Beijing Engineering Research Center of High-speed Railway Broadband Mobile Communications, and the School of Electronic and Information Engineering, Beijing Jiaotong University, Beijing 100044, China (e-mail:
niuy11@163.com).
}

\thanks{Y. Liu and Y. Li are with State Key Laboratory on
 Microwave and Digital Communications, Tsinghua National Laboratory for Information
 Science and Technology (TNLIST), Department of Electronic Engineering, Tsinghua
 University, Beijing 100084, China (e-mail: liyong07@tsinghua.edu.cn).}



\thanks{
 P. Hui is with the Department of Computer Science and Engineering, the Hong Kong University of Science and Technology, Hong Kong.}
\thanks{This study was supported by the National Natural Science Foundation of China Grants 61725101 and 61801016; and by the China Postdoctoral Science Foundation
under Grant 2017M610040 and 2018T110041; and by National key research and development program under Grant 2016YFE0200900; and by
the Beijing Natural Fund under Grant L172020;
and by Major projects of Beijing Municipal Science and Technology Commission under Grant No. Z181100003218010.}
}

\maketitle

\begin{abstract}

As an extension of cloud computing, fog computing at the edge of networks provides low latency, location awareness, and real-time interactions.
At the same time, millimeter wave (mmWave) communications are able to provide directional multi-gigabit transmission rates with large available bandwidth.
Based on the user mobile trajectories in a region, several activity hotspots that users pass by frequently can be obtained.
By caching popular content at the edge nodes near the hotspots, users can download the cached content directly at a short distance, and
the user experience can be significantly improved. Considering multiple hotspots in a region, how to efficiently schedule the transmission for the caching at edge nodes
becomes a key problem. In this paper, we focus on the problem of mobility aware transmission scheduling for caching at edge nodes near hotspots, and
utilize multi-hop relaying and concurrent transmissions to achieve better performance.
After formulating the optimal scheduling problem as a stochastic nonlinear mixed integer program,
we propose a mobility aware caching scheduling scheme, called MHRC (Multi-Hop Relaying based Caching), where multi-hop D2D paths are established for edge nodes,
and concurrent transmissions are exploited in the scheduling of caching at edge nodes.
Extensive performance evaluation demonstrates MHRC achieves more than 1x higher expected cached data amount compared with state-of-the-art schemes.

\end{abstract}

\section{Introduction}\label{S1}

Recently, fog computing has received much attention from both academia and industry
as an extension of cloud computing, and provides
low latency, location awareness, real-time interactions, and mobility support \cite{fog}.
In fog computing, computation, storage, and networking services are provided between end devices and data centers located at the edge of network.
Thus, compute-intensive services can be offloaded to the edge of networks to save the backhaul bandwidth and provide better services for users.
Therefore, there is also similar concept known as edge computing.
On the other hand, there is huge bandwidth available in the millimeter wave (mmWave) band, and multi-gigabit broadband communication services can be provided by mmWave communications in the fifth generation (5G) era \cite{mmWave1,mmWave2,mmWave3}. To overcome high propagation loss in the mmWave band, directional antennas are utilized to achieve high antenna gain.
Consequently, concurrent transmissions are exploited to improve mmWave network capacity \cite{concurrent1,concurrent2,concurrent3}.
Thus, mmWave communications become an effective way to realize low latency and high data rate communication in fog computing systems.

\textbf{Motivations:}
To realize fog computing, advanced techniques such as multi-hop D2D relaying transmissions should be leveraged \cite{multihop, CachinMobile}.
By establishing multi-hop D2D transmission paths to the edge nodes, the advantages of D2D communications such as saving power and higher spectral efficiency can be reaped to provide better services for users \cite{D2D-Doppler,zhao2015social}.
In \cite{concurrent3}, multi-hop D2D transmissions are exploited to improve the flow throughput.
On the other hand, due to weak diffraction ability, mmWave links are sensitive to blockages.
Exploiting multi-hop D2D communications can also improve link connection robustness since the length of each D2D link is shorter as in \cite{concurrent3}.
In a region, there are several activity hotspots where users including pedestrians or vehicles pass by frequently, such as restaurants, stadiums, tourist attractions, and cinemas.
During staying at the hotspots, popular content or data can be cached near the hotspots to accommodate the mobile traffic demands from users.
Through the multi-hop D2D transmission paths, popular content for users can be cached via high data rate mmWave communication at the edges nodes,
and when users pass by this hotspot, the cached content can be downloaded by users from the edge nodes instead of from the base station.
The edge nodes in our system are deployed near the hotspots, and better user experiences can be achieved through better channel conditions between edge nodes and users.
By caching popular content at edge nodes in advance,
users can obtain more content and user experience can be improved significantly.

Considering the user mobile trajectories in the region, the hotspots will have different ``hot'' levels, i.e., users will pass by some hotspots more
frequently, while less frequently for some hotspots. How to cache data near the hotspots to maximize the user experience becomes a key problem, and
the statistical properties of user mobility should be considered in the solution.


In this paper, we propose to exploit the multi-hop D2D relaying in the mmWave band to cache content for mobile users in a region consisting of multiple activity hotspots. The base stations host the data centers at the edge of the network.
Furthermore, we deploy relay nodes from the base station to the hotspots, and multi-hop D2D relaying in the mmWave band is exploited to improve the expected total amount of data received during their passing by the hotspots. The relay nodes deployed near the hotspots are called edge nodes in the system.
The data content is downloaded from the base station via the multi-hop D2D relaying of relay nodes and cached at the edge nodes.
In a fixed period, the transmissions for caching at the edge nodes are scheduled, and afterwards users can receive cached data from the edge node when passing by the hotspots.
Considering the mobile users pass by the region through different hotspots, we try to maximize the expected amount of data received by users during passing by.
Our novel contributions are summarized as follows.
 \begin{itemize}
 \item We formulate the problem of optimal mobility aware multi-hop D2D relaying in the mmWave band for caching at edge nodes of hotspots as a stochastic nonlinear mixed integer programming problem to maximize the expected amount of data received during users passing by. Under directional transmissions, concurrent transmissions are exploited to improve efficiency. The probabilities that users pass by the hotspots are considered into the problem.
 \item We propose a multi-hop D2D relaying based caching scheme, called MHRC (Multi-Hop Relaying based Caching), which establishes the multi-hop D2D transmission paths for edge nodes near the hotspots and schedules the transmissions for caching at the edge nodes. The probabilities that users pass by the hotspots obtained from user mobile trajectories are explicitly considered in MHRC.

 \item Extensive evaluations on MHRC algorithm under various system parameters are done. MHRC achieves signi?cantly better performance in terms of the expected cached data amount compared with other existing schemes.
     With the algorithm parameters optimized, MHRC is able to achieve more than 1x higher expected cached data amount.

\end{itemize}

The rest of the paper is organized as follows. In section \ref{sec:RW}, related works on directional protocols for WPANs or WLANs in the mmWave band are discussed.
Section \ref{sec:System-Model} introduces the system model, and formulates the optimal problem of maximizing the expected amount of data received during users passing by the hotspots as a stochastic nonlinear mixed integer programming problem. Section \ref{MHRC} presents our proposed MHRC scheme. Section \ref{sec:Simulation} presents the performance evaluation under various system parameters.
In the light of our results, we conclude this paper in Section  \ref{sec:Conclusion}.
\\

\section{\label{sec:RW}Related Works}


There are some related works on content caching and distribution in fog systems. Many works consider this aspect in the context of
Fog-Radio Access Networks (F-RAN) \cite{fog97, fog98, fog99, fog100, fog101}. F-RAN extends the concept of C-RAN, and complements it
with part of radio functionalities close to users. Content can be cached near users to serve users with low latency.
Tandon \emph{et al. } \cite{fog97} analyze the performance of the F-RAN system focusing on the trade-off among latency, caching, and fronthaul capacity metrics.
Park \emph{et al. } \cite{fog98} analyze the performance of hard and soft transfer modes in the F-RAN system to serve content to users.
Hung \emph{et al.} \cite{fog99} analyze which content to cache in the cloud and which content to cache in the fog stratum based on the content features.
Xiang \emph{et al.} \cite{fog100} consider that the content can be cached over the user equipment. Then they maximize the energy efficiency by mode
selection and resource allocation in serving content to users. Three different communication modes are considered, the device to device model, single serving
antenna, and coordination among antennas.
Do \emph{et al.} \cite{fog101} consider the management of content in a fog-based content distribution network to optimize the content distribution from the data centers
to the fog nodes. The authors propose a distributed proximal algorithm to solve the problem with a small number of iterations.
However, none of the schemes above exploits the Gbps mmWave communication in the multi-hop D2D transmission for efficient caching scheduling
in fog computing systems.

There are already some works on directional protocols for WPANs or WLANs in the mmWave band \cite{concurrent1,concurrent2,concurrent3,Qiao,MRDMAC,mao,EX_Region,Qiao_7,JSAC_own,tvt_own,chenqian,G_theta,singh_outdoor}.
In \cite{EX_Region}, Cai \emph{et al.} proposed the
REX scheduling scheme (REX) based on the exclusive region
(ER) conditions that concurrent
transmissions always outperform time division multiple access (TDMA).
Qiao \emph{et al.} \cite{Qiao} proposed a concurrent transmission scheduling algorithm to maximize the number of flows with the throughput requirement of each flow satisfied.
Furthermore, a multi-hop concurrent transmission scheme is proposed to address the link outage
problem (blockage) and combat huge path loss to improve flow throughput \cite{Qiao_7}.
Son \emph{et al.}
\cite{mao} proposed a greedy coloring algorithm in the frame based directional MAC protocol (FDMAC) to exploit concurrent transmissions for high efficiency. Singh \emph{et al.} \cite{MRDMAC} exploited multi-hop relay to overcome blockage in the multi-hop relay directional MAC protocol (MRDMAC).
In \cite{JSAC_own}, Niu \emph{et al.} proposed a joint scheduling scheme for the radio access and
backhaul of small cells in the mmWave band, where D2D communications are exploited to improve the system performance.
In \cite{tvt_own}, two-hop relaying is adopted to overcome the blockage problem in the mmWave WPANs, and concurrent transmissions and relay selection are optimized jointly to improve the network throughput.
In \cite{chenqian}, Chen \emph{et al.} proposed a directional cooperative MAC protocol to coordinate the
uplink channel access of an IEEE 802.11ad WLAN. Based on IEEE 802.11 ad, Chen \emph{et al.} \cite{G_theta} proposed a spatial reuse
strategy to schedule two different SPs to overlap with each other, and also analyzed the
performance of the strategy with the difference between idealistic and realistic directional
antennas considered.

Some works are on utilizing the mmWave communications for wireless backhauling. In \cite{Green}, an energy efficient scheduling scheme was proposed for the backhauling of small cells in the mmWave band, where concurrent transmissions are exploited to reduce energy consumption via power control.
To control the interference between links, the contention graph and maximum independent set are utilized to fully unleash the potential of concurrent transmissions.
In \cite{Wu1}, Zhu \emph{et al.} proposed the Regret Benefit Ratio Scheduler (RBRS) that is able to maximize the number of links with their QoS requirements satisfied. Furthermore, we also propose a distributed scheme under CSMA/CA, where the regret benefit ratio (RBR)
is implemented by prioritizing MAC contention window. In \cite{Wu2}, the distributed maximum QoS-aware (DMQ)
scheduling algorithm is proposed for the mmWave backhaul network of small cells to maximize the system throughput
while satisfying QoS requirements for each flow.

With large bandwidth available in the mmWave band, Gbps data rates can be supported in the mmWave band, and
mmWave communications can achieve efficient transmission for caching in fog systems. Furthermore, concurrent transmissions
can be exploited to improve caching efficiency due to less interference in the mmWave band.
To the best of our knowledge, we are the first to exploit multi-hop D2D communications and concurrent transmissions in the mmWave band to improve
the caching efficiency in the fog computing system.

\section{\label{sec:System-Model}System Model and Problem Formulation}

\subsection{System Model}


\begin{figure}[htbp]
\vspace*{-1mm}
\begin{minipage}[t]{1\linewidth}
\centering
\includegraphics[width=0.9\columnwidth]{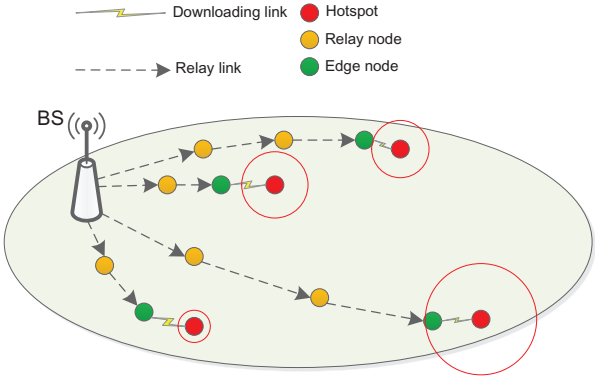}
\end{minipage}%
\vspace*{-3mm}
\caption{Multi-hop D2D caching for hotspots in fog computing systems.}
\label{fig:fogcomputing}
\vspace*{-1mm}
\end{figure}

In Fig. \ref{fig:fogcomputing}, we illustrate the multi-hop D2D caching at edge nodes for hotspots in fog computing systems. Fog extends the central cloud from the data center to the edge of networks.
The fog consists of the base station (BS) and some relay nodes (REs). The relay nodes near the hotspots are called the edge nodes.
The base station (BS) communicates with the edge nodes through the multi-hop relaying via the relay nodes in the mmWave band.


There is one base station in the system, and several hotspots in the service region. The base station is denoted by $B_s$.
System time is partitioned into time slots of
equal length, and $K$ time slots are allocated for content caching at edge nodes near hotspots. The caching transmission is scheduled by the base station. For each hotspot, a multi-hop D2D relay path from the base station to the edge node is established, and the content is cached through the multi-hop transmission from the base station to the edge node.
We denote the set of relay nodes including the edge nodes in the system by $V$. The set of hotspots is denoted by $\mathbb{U}$. For each user, which may be a pedestrian or a car, the set of hotspots it passes by is denoted by $\mathbb{U}_0$. We know $\mathbb{U}_0  \subseteq  \mathbb{U}$. For each hotspot $u\in \mathbb{U}$, we denote its edge node by $s_u$, and the user stays at the hotspot for $\tau_u$ time slots.

 The system time is partitioned into non-overlapping time slots of equal length, and the BS synchronizes the clocks of REs. Equipped with electronically steerable directional antennas, directional transmissions are supported between any pair of BS and REs. Besides, we assume all nodes are half-duplex, i.e., each node has at most one connection with another node simultaneously. Multi-hop D2D paths are established for each hotspot, and we call the nodes on the paths by D2D relays, which are shown as the orange circles in the figure. When users pass by the hotspots in this region, users can download the cached data from the edge nodes (green circles) near the hotspots (red circles). We also use the red circles outside the hotspots to represent how hot the hotspot is, i.e., the probability that users will pass by the hotspot. Larger circles means larger probability.

The transmissions for caching at the edge nodes near the hotspots are scheduled in a frame consisting of finite time slots.
In the system, the BS makes the scheduling decisions for caching at the edge nodes, and pushes the scheduling decisions to the relay nodes to establish the multi-hop D2D paths. With the schedule obtained, the relay nodes start transmissions for caching at the edge nodes according to the schedule. The scheduling decisions will be updated periodically considering the different patterns of users passing by the region in different periods like weekends and weekdays.

We denote the directional link from node $i$ to $j$ by $(i,j)$. Then according to the path loss model, the received power at node $j$ for $(i,j)$, denoted by ${{P^r_{ij}}}$ (mW), can be obtained as
\begin{equation}
{P^r_{ij}} = {k_0}{G_t}(i,j){G_r}(i,j){l_{ij}^{ - \gamma }}{P_t},
\end{equation}
where $P_t$ (mW) is the transmission power, $k_0$ is a constant coefficient and proportional to ${(\frac{\lambda }{{4\pi }})^2}$ ($\lambda $ denotes the wavelength), ${G_t}(i,j)$ is the transmit antenna gain of node $i$ in the direction from node $i$ to $j$, and ${G_r}(i,j)$ is the receive antenna gain in the direction from node $i$ to $j$, $l_{ij}$ (m) is the distance from transmitter $i$ to receiver $j$, and $\gamma$ is the path loss exponent \cite{Qiao}.

Since directional transmissions, there is less interference between links, and concurrent transmissions of links can be exploited to improve network capacity \cite{Qiao}. If link $(u,v)$ and link $(i,j)$ are scheduled concurrently, the interference power at node $j$ from node $u$ can be obtained as
\begin{equation}
I_{uvij} =\rho k_0{G_t}(u,j){G_r}(u,j){l_{uj}}^{ - \gamma }{P_t},
\end{equation}
where $\rho$ is related to the cross correlation of signals from different links \cite{Qiao}.

We denote the set of links scheduled concurrently with link $(i,j)$ by $\mathbb{C}_{ij}$. Then the received signal to interference plus noise ratio (SINR) at receiver $j$ can be expressed as
\begin{equation}
{\Gamma_{ij}} = \frac{{{k_0}{G_t}(i,j){G_r}(i,j){l_{ij}}^{ - \gamma }P_t}}{{{N_0}W + \rho \sum\limits_{(u,v) \in {\mathbb{C}_{ij}}} {k_0{G_t}(u,j){G_r}(u,j){l_{uj}}^{ - \gamma }{P_t}} }},
\end{equation}
where $W$ (Hz) is the bandwidth  and $N_0$ (mW/Hz) is the one-sided power spectra density of white Gaussian noise \cite{Qiao}. Considering the reduction of multipath effect for directional mmWave links \cite{MRDMAC}, the achievable data rate for link $(i,j)$ can be obtained according to Shannon's channel capacity as
\begin{equation}
\begin{array}{l}
{R_{ij}} = \eta W{\log _2}\left(1 + \frac{{{k_0}{G_t}(i,j){G_r}(i,j){l_{ij}}^{ - \gamma }P_t}}{{{N_0}W + \rho \hspace{-0.4cm}\sum\limits_{(u,v) \in {\mathbb{C}_{ij}}}\hspace{-0.4cm} {k_0{G_t}(u,j){G_r}(u,j){l_{uj}}^{ - \gamma }{P_t}} }}\right),
\end{array}\label{rate}
\end{equation}
where $\eta \in (0,1)$ depicts the efficiency of the transceiver design \cite{Qiao}.

\subsection{Problem Formulation}

We define a binary variable $b^k_{ij}$ to indicate whether link $(i,j)$ is scheduled in the $k$th time slot. Then we can obtain the transmission rate of link $(i,j)$ at the $k$th time slot as
\begin{equation}
\begin{array}{l}
{R_{ij}^k} = \eta W{\log _2}\left(1 + \frac{{b^k_{ij}{k_0}{G_t}(i,j){G_r}(i,j){l_{ij}}^{ - \gamma }P_t}}{{{N_0}W + \rho \sum\limits_{(u,v) } {b^k_{uv}k_0{G_t}(u,j){G_r}(u,j){l_{uj}}^{ - \gamma }{P_t}} }}\right),
\end{array}\label{rate2}
\end{equation}

For each hotspot $u\in \mathbb{U}$, the number of hops of the relay path for $u$ is denoted by $H_u$, and the receiver of the $j$th hop is denoted by $w_j(u)$. The duration of each time slot is denoted by $\Delta$.
For the $j$th hop, the maximum amount of data cached at $w_j(u)$ can be obtained as

\begin{equation}
D_{w_j(u)}=\sum\limits_{k = 1}^K {R_{{w_{j-1}(u)}{w_j(u)}}^k\Delta }.
\end{equation}

For the hotspot $u$, we denote the amount of data the user received by $D_u$. Then for a particular user, the total amount of data received during its passing by this region can be obtained as $\sum\limits_{u \in \mathbb{U}_0} {{D_u}} $. Then for multiple users, the expected total amount of data received during their passing by this region can be obtained as $E(\sum\limits_{u \in {\mathbb{U}_0}} {{D_u}} )$, where $E( \cdot )$ denotes the expectation operation.

To optimize the user service, the amount of cached data received during its passing by should be maximized. Therefore, the objective function to be maximized is $E(\sum\limits_{u \in {\mathbb{U}_0}} {{D_u}} )$.
We now consider the system constraints. First, due to the half-duplex assumption, each node has at most one connection with one neighbor, i.e., adjacent links cannot be scheduled concurrently, which means that,
\begin{equation}
\begin{aligned}
b_{uv}^k + b_{ij}^k \leq  1, &{\rm{if \;link}}\; (u,v)\; {\rm{and}}\; (i,j) {\rm{\ are\; adjacent}},\\ & \forall\; u,v,i,j \in  V \cup B_s. \label{cons1}
\end{aligned}
\end{equation}

Second, due to inherent transmission order, the preceding hops on the relay path for each hotspot should be scheduled ahead of hops behind, which can be expressed as,

\begin{equation}\label{cons2}
\begin{aligned}
 \sum\limits_{k=1}^{K^*} b_{{w_{j-2}(u)}{w_{j-1}(u)}}^k \ge \sum\limits_{k=1}^{K^*} b_{{w_{j-1}(u)}{w_j(u)}}^k , \ \ \\&\hspace{-3cm}
  \forall j\ge 2, u\in \mathbb{U}_0, \ K^*\in\{1,\cdots, K\},
\end{aligned}
\end{equation}
where $K^*$ varies from $1$ to $K$.

Third, the amount of data the user receives at the hotspot cannot exceed the maximum amount of data it can receive, i.e.,
\begin{equation}\label{cons3}
 D_u \le R_{{s_u}u}{\tau _u},  \forall u\in \mathbb{U}_0,
\end{equation}
where $R_{{s_u}u}$ is the transmission rate from the edge node to the hotspot $u$.

Fourth, the amount of data the user receives at the hotspot is determined by the amount of cached data by nodes on the relay path, and thus
$D_u$ cannot exceed the maximum amount of data that can be cached by each hop on the relay path, which can be expressed as

\begin{equation}\label{cons4}
D_u \le D_{w_j(u)},    1\le j \le H_u, \forall u\in \mathbb{U}_0.
\end{equation}

Therefore, the optimal multi-hop D2D relay caching problem for users (P1) can be formulated as follows.

\begin{equation}\hspace{-2.3cm}
({\rm{P1}})\ \ \max E(\sum\limits_{u \in {\mathbb{U}_0}} {{D_u}} ), \label{obj2}
\end{equation}
\hspace{2.50cm}s. t.
\hspace{0.2cm}Constraints (\ref{cons1})--(\ref{cons4}).
\\

In Problem P1, the parameters for users passing by the region, $\mathbb{U}_0$ and $\tau _u$, are random variables. Besides, there are also nonlinear terms
in constraints (\ref{cons3}) and (\ref{cons4}) due to the equation (\ref{rate2}) for transmission rate. We can observe that $D_u$ is a real variable, and $b_{ij}^k$ is a binary variable.
Therefore, problem P1 is a stochastic nonlinear mixed integer programming problem, which is NP-hard \cite{mao,NP}.
Obtaining the optimal solution requires full information regarding the statistical properties of user mobility trajectories, and the exhaustive
search is needed generally, which requires much computation time.
In the next section, we propose a heuristic multi-hop D2D relay caching scheme to achieve excellent performance with low complexity.

\section{\label{MHRC}The MHRC Scheme}

In this section, we propose a mobility aware relaying based caching scheme. First, we introduce the relay path planning algorithm
to establish the multi-hop relay paths for edge nodes near the hotspots. Then, the caching scheduling algorithm is presented to schedule the transmissions for caching at the edge nodes near the hotspots. In the scheduling algorithm, the probabilities that users pass by the hotspots
obtained from the user mobile trajectories are explicitly considered, and the hotspots with higher probabilities are scheduled first.
Finally, the data updating algorithm is proposed to obtain the
actual data amount cached at the edge nodes.

\subsection{Mobility-aware Relay Path Planning}

For each hotspot $u$, we first choose its nearest RE as the edge node $s_u$ for the best channel conditions. Then we establish a multi-hop transmission path from the BS to $s_u$ with predetermined hop number, borrowing the design ideas of Dijkstra Algorithm \cite{Dijkstra}. If the found path's hop number does not equal to the predetermined one, we remove the path, and find a new path with the right hop number.

We denote the BS in the network by $B_s$, and the edge node $s_u$ by $E$. The set of selected paths is denoted by $\mathbb{P}$, and the hop number for each path $p\in \mathbb{P}$ is denoted by $H_p$. For each path $p\in \mathbb{P}$, we denote the last node on $p$ by $L_p$. Besides, we denote the set of visited nodes in planning by $\mathbb{N}_V$.
The selected path from $B_s$ to $E$ is denoted by $p_*$, and the predetermined hop number for $p_*$ is denoted by $H_*$.
We also denote the minimum distance from $E$ to other nodes in $\mathbb{V}$ by $l_{EV}$, to indicate whether there exists the path to $E$. If $l_{EV}$ equals to infinity, no path exists from $B_s$ to $E$ with $H_*$ hops.

\begin{algorithm}[htbp]
\caption{Mobility-aware Relay Path Planning Algorithm}\label{algorithm1}
\textbf{Input:} The set of REs in the cell, $\mathbb{V}$;\\
\qquad \quad The predetermined hop number for $p_*$, $H_*$;\\
\textbf{Function} $p_*$=\textbf{PathPlanning} ($\mathbb{V}$, $H_*$, $E$)\\
\textbf{Initialization:}
$\mathbb{N}_V=\{B_s\}$, $\mathbb{P}=\emptyset$, $p_*=\emptyset$, $l_{EV}= 0$, $\mathbb{V}_{ini}=\mathbb{V}$;\\
\While{$E \notin \mathbb{N}_V$ $\rm{and}$ $l_{EV}\neq \infty$}
{\For{\rm{each node} $i \in\mathbb{N}_V$}
{Find the relay node $v \in \mathbb{V}$ with the shortest distance to node $i$;\\
$r_{i}=l_{vi}$, $c_i=v$;
}
Find the node $s \in \mathbb{N}_V$ with the minimum $r_s$;\\
\If{$s== B_s$}
{
$p={\{B_s\to{c_s}\}}$;\\
$\mathbb{P}=\mathbb{P}\cup{p}$;
}
\Else
{Find the path $p_0\in{\mathbb{P}}$ with $L_{p_0}=s$;\\
Obtain $p$ by extending $p_0$ to $c_s$, $p=\{p_0\to{c_s}\}$;\\
$\mathbb{P}=\mathbb{P}\cup{p}$;
}
\If{$c_s== E$}
{\If{$H_p==H_*$}
{$p_*=p$, $\mathbb{N}_V=\mathbb{N}_V \cup{E}$;}
\Else
{$\mathbb{P}=\emptyset$;
$l_{Es}=\infty$; $\mathbb{N}_V=\{B_s\}$; $\mathbb{V}=\mathbb{V}_{ini}$;}
}
\Else
{$\mathbb{N}_V=\mathbb{N}_V \cup{c_s}$, $\mathbb{V}=\mathbb{V}-c_s$;}
Update the shortest distance from $E$ to other nodes in $\mathbb{V}, l_{EV}$;
}
$\mathbf{Return}\ p_*$.
\end{algorithm}

The pseudo-code of the mobility-aware relay path planning algorithm is presented in Algorithm \ref{algorithm1}. The algorithm can be represented as a function \textbf{PathPlanning}, as in line 3. The algorithm iteratively searches for the relay path with predetermined hop number $H_*$ until the path is found or there is no path that can be found, as indicated in line 5. Then the algorithm extends to find the next node with the minimum distance to one of the visited nodes, as in lines 6--9. If the visited node is $B_s$, a new path from $B_s$ to the next node is generated, and included in $\mathbb{P}$, which is indicated by lines 10--12. If the visited node is not $B_s$, the algorithm finds the corresponding path that the visited node is on,
and extends the path to the next node, as shown in lines 14--16. If the newly added or extended path goes to the edge node $E$, the algorithm examines the hop number of the path, and if the hop number is $H_*$, the relay path $p_*$ is obtained and returned as in lines 17--19. Otherwise, the algorithm sets the distance of the last hop on $p$, $l_{Es}$, to infinity, which disables the link being on the relay path. Then the algorithm will search for the possible path again after setting $\mathbb{N}_V$, $\mathbb{V}$, and $\mathbb{P}$ as initial values. If the newly added or extended path does not go to the edge node, the next node will be added to $\mathbb{N}_V$, and removed from $\mathbb{V}$. The minimum distance from $E$ to other nodes in $\mathbb{V}$, $l_{EV}$, is updated in line 24. Line 25 returns the obtained D2D path $p_*$.


From Algorithm \ref{algorithm1}, we can infer that the outer while loop has no more than $2|\mathbb{V}|$ iterations, and the inner for loop has at most ${|\mathbb{V}|}$ iterations. Consequently, the worst case computational complexity of Algorithm \ref{algorithm1} is $\mathcal{O}(|\mathbb{V}|^2)$, which is achievable in practice.

\subsection{Caching Scheduling}

After establishing the relay paths from the BS to the edge nodes for the hotspots by the relay path planning algorithm, we propose a caching scheduling algorithm to schedule the transmissions for caching at the edge nodes.

%
%
%
\subsubsection{Concurrent Transmission Conditions}

To improve system performance, concurrent transmissions are exploited in caching scheduling. On the other hand, interference between links
should be controlled. To enable concurrent transmissions, we require the interference each concurrent link suffers should be less than a predetermined value. Concretely, for a concurrent link set $V$, the following condition should be met before link $(i,j)$ is added to $V$.
\begin{equation}
\sum\limits_{(u,v) \in V \cup \{ (i,j)\} } {{I_{uvxy}}}  < \sigma  \cdot {P_t},\;{\rm{for}}\;{\rm{each}}\;(x,y) \in V \cup \{ (i,j)\},
\end{equation}
where we use $\sigma$ to characterize the interference threshold. In this way, the interference among concurrent links is controlled.
In the following, if one link cannot be added into one concurrent link set for concurrent transmissions, i.e., the concurrent transmission conditions are not satisfied,
we denote that this link has contention with links in the set.
Due to the half-duplex assumption, adjacent links will always have contention between each other, and thus cannot be scheduled concurrently as in constraint (\ref{cons1}).

\subsubsection{Caching Scheduling Algorithm}

The caching scheduling algorithm iteratively schedules the transmissions for links on the relay paths from the BS to the edge nodes into each time slot. To maximize the amount of cached data, the algorithm tries to schedule the transmissions with as much data as possible. If the cached data amount cannot be scheduled successfully within $K$ time slots, it will be reduced by multiplying a shrinking factor, which is smaller than 1. Besides, the algorithm tries to establish relay paths for edge nodes with as more hops as possible, and if relay paths with more hops cannot be scheduled successfully, the hop number will decrease. In each time slot, concurrent transmissions of links are enabled if the concurrent transmission conditions are satisfied. The algorithm schedules the transmissions for caching at edge nodes in decreasing order of the probabilities that users pass by the hotspots, and the algorithm terminates until all edge nodes for hotspots are examined.

First, we introduce the notations in the caching scheduling algorithm.
For each hotspot $u\in \mathbb{U}$, we denote the probability that users pass by as $f(u)$, and the probability vector is denoted by $\boldsymbol{f}$.
The probabilities for users are obtained via statistics methods from a large amount of user mobility trajectories in the region.
Considering the probabilities are different in different days, e.g., the probabilities are different on weekdays and weekends, the probabilities will be updated every single day.
We can also obtain the expected stay time of users at each hotspot, and for each hotspot $u \in \mathbb{U}$, the expected stay time is denoted by
$\tau_u$, and the expected stay time vector of users at hotspots is denoted by $\boldsymbol{\tau}$.
The maximum hop number for the relay paths from the BS to the edge nodes is denoted by $H_m$. The hop number for the relay path from the BS to $s_u$
is denoted by $H_u$. For the edge node $s_u$, the relay path from the BS to $s_u$ is denoted by $p_u$. The $h$th hop link on $p_u$ is denoted by ${p_u^h}$. For ${p_u^h}$, the last time slot it occupies for transmission in the schedule is denoted by $k_u^h$.
The start time slot of ${p_u^h}$ for transmission in the algorithm is denoted by $s_u^h$, and if $h=1$, $s_u^h$ is equal to $k_{u_l}^1+1$, where
$k_{u_l}^1$ is the last time slot the previous scheduled link ${p_{u_l}^1}$ occupies for transmission. If $h>1$, $s_u^h$ is equal to $k_u^{h-1}+1$.
For each hotspot $u$, the number of remaining available time slots is denoted by $r_u$. Correspondingly, the vector for $\mathbb{U}$ is denoted by $\boldsymbol{r}$. The estimated amount of data cached at the edge node for hotspot $u$ is denoted by $D''_u$, and the corresponding vector is denoted by $\boldsymbol{D}$. In the caching scheduling algorithm, $K$ time slots are allocated for caching scheduling.
In the $k$th time slot, the set of links scheduled for transmission in this time slot is denoted by $V^k$.

\begin{algorithm}\label{algorithm2}
\caption{Caching Scheduling Algorithm}
\textbf{Input:} The set of hotspots $\mathbb{U}$, the set of REs $\mathbb{V}$;
 Probability vector $\boldsymbol{f}$ for $\mathbb{U}$;
 Expected stay time vector $\boldsymbol{\tau}$ for $\mathbb{U}$;\\
\hspace{9.5mm} The maximum hop number $H_m$; Interference threshold $\sigma$, the shrinking factor $\beta$;\\

\textbf{Initialization:} $\boldsymbol{r}=\textbf{0}$, $\boldsymbol{D}=\textbf{0}$, $k$=0, $\mathbb{P}=\emptyset$; Set $V^k=\emptyset$; \\

\While{$|\mathbb{U}|>0$}
{Find the edge node $s_u$ with the maximum $f(u)$;\\
 $D''_u=R_{s_{u}u}\cdot{{\tau}_{u}}$;\\ $H_u=H_m$;\\
\While{$H_u > 1$}
{$p_u$=$\textbf{PathPlanning}$ ($\mathbb{V}$, $H_u$, $s_u$);\\
\If{$p_u \neq{\emptyset}$}
{
Obtain the rate of $p_u^i$ with the maximum possible interference, $R_i^u$;\\%
$r_u=K-\sum_{i=1}^{H_u}{\frac{D''_u}{R_i^u}}-k_{u_l}^1$;   $h=1$;\\
\If{$r_u\ge0$}
{
\While{$h\le H_u$}
{Obtain the start time slot for $p_u^h$, $s_u^h$;\\
Obtain the needed time slots for $p_u^h$, $T_u^h=\frac{D''_u}{R_h^u}$;\\
Set $k'$=$s_u^h$, $T_{d}=0$;\\
\While { $k'\leq K$, $T_d<T_u^h$, \rm{and} $r_u\ge0$ }
{Set $V^{k'}_t=V^{k'}$;\\
Check whether $p_u^h$ \rm{has contention with links in} $V^{k'}_t$;\\
\If{\rm{FALSE}}
{
$V^{k'}_t=V^{k'}_t\cup p_u^h$;\\
$T_d=T_d+1$;
$k'=k'+1$;
}
\Else{ $r_u=r_u-1$;$k'=k'+1;$}
}
\If{$r_u \geq 0$ \rm{and} $k'\leq K$}
{
Go to line 36 for the next hop;}
\Else{$H_u=H_u-1$;\\Go to line 13 for a new path;}
$h=h+1$;\\
}
Update all $V^k$ by $V^k_t$;
$\mathbb{P}=\mathbb{P}\cup p_u$;\\
}}
\Else{$H_u=H_u-1$;}
}
\If{$H_u==1$}
{$D''_u=\lfloor D''_u\cdot{\beta}\rfloor$;\\
\If{$D''_u==0$}
{Go to line 46 for the next hotspot;}
Go to line 12 with less caching data;
}
  $u_l=u$;\\  $\mathbb{U}=\mathbb{U}-u$; \\
}
$\mathbf{Return}$ $V^k$ for each time slot, and $\mathbb{P}$ for relay paths.

\end{algorithm}


The pseudo-code of caching scheduling algorithm is presented in Algorithm \ref{algorithm2}.
First, the set of links in each time slot is initialized to an empty set, as shown in line 3.
The algorithm schedules the transmissions for caching at each edge node in decreasing order of the probabilities that users pass by the hotspots until all the edge nodes are examined, as shown in lines 4--5. In line 6, the expected maximum amount of data that can be received by users passing by the hotspot is obtained, which is also the initial value of the amount of data cached at the edge node $D''_u$.
In line 7, the hop number for the edge node of hotspot $u$ is initialized to the maximum hop number, $H_m$.
In lines 8--34, the algorithm establishes relay paths with as more hops as possible, and schedules the transmissions for links on the relay path into appropriate time slots.

\textbf{Relay path generation}:
In line 9, the path planning function generates the relay path of $H_u$ hops from the BS to the edge node $s_u$ if $H_u$ is larger than 1. Otherwise, $D''_u$ decreases by multiplying a shrinking factor $\beta$ as in lines 35--36, and the algorithm tries to establish relay paths with the new $D''_u$ as in line 39. If $D''_u$ decreases to 0, which indicates the scheduling for caching at this edge node is failed, the algorithm will start to schedule the transmission for caching at the next edge node of the corresponding hotspot as in lines 37--38.
If the generated path is not empty, the algorithm calculates the transmission rate of links on the path with the maximum possible interference considered, i.e., the interference is equal to $\sigma  \cdot P_t$.
Then the algorithm obtains the remaining number of time slots for $s_u$, $r_u$. If the generated path is empty, $H_u$ decreases by 1, and a new path will be generated.
 If $r_u \ge 0$ in line 13, which indicates transmissions on the relay path may be scheduled successfully
by the algorithm, the algorithm starts the scheduling for links on the relay path from previous hops to hops behind, as in lines 14--31.
In line 15 and 16, the algorithm obtains the start time slot and the number of needed time slots for $p_u^h$, $s_u^h$ and $T_u^h$.
Then in line 17, the algorithm tries to schedule the transmission for the $h$ hop link on $p_u$, $p_u^h$, and $T_d$ is defined to record the number of successfully scheduled time slots for $p_u^h$. As in line 18, the algorithm iteratively schedules the transmission for $p_u^h$ into multiple time slots until enough time slots are scheduled
or there are not enough time slots for $p_u^h$. In line 19, $V^{k'}_t$ is defined to record the temporary value of $V^{k'}$.
In line 20, the algorithm checks whether $p_u^h$ has contention with links already in this time slot.
If there is no contention, $p_u^h$ will be added to this time slot in line 22; otherwise, $p_u^h$ will not be scheduled into this time slot, and the next time slot will be examined in line 25. In line 26, when the transmission for $p_u^h$ has been scheduled successfully, the algorithm executes
line 31 for the scheduling of the next hop. If the scheduling for this hop fails, the hop number $H_u$ decreases, and a new path with fewer hops is generated as in lines 29--30. After the scheduling for hops on $p_u$, all $V^{k}$ are updated by $V^{k}_t$, and the generated relay path is included into $\mathbb{P}$ as in line 32.

\textbf{Scheduling results output}:
In line 40, the hotspot whose caching transmission has been scheduled successfully is recorded by $u_l$ for the calculation of $r_u$ in line 12 and $s_u^h$ in line 15. The examined hotspot is removed from $\mathbb{U}$ in line 41. In line 42, the scheduling result for each time slot $V^{k}$ and the set of generated relay paths for hotspots $\mathbb{P}$ are returned.

For Algorithm \ref{algorithm2}, we can observe that the while loop in line 4 has no more than $|\mathbb{U}|$ iterations, and the while loop in line 8 has at most $H_m$ iterations. Considering the complexity of Algorithm \ref{algorithm1},
the worst case computational complexity for Algorithm \ref{algorithm2} is $\mathcal{O}({H_m}|\mathbb{U}||\mathbb{V}|^2)$, which can be implemented in practice.

\subsubsection{Data Updating Algorithm}

Due to the factual interference, the transmission rates of links can be improved, and the actual data amount that can be cached should be updated. In the following, we propose a data updating algorithm to calculate the actual data amount cached at each edge node.
For the memory at each edge node, the amount of data that can be cached at each edge node is limited by the transmission rate of the mmWave link between the edge node and the user, and the stay time of the user. Besides, the amount of data that can be cached is also limited by the limited number of time slots for scheduling. Thus, we assume the memory at each edge node is enough to cache the data.

The amount of data cached at the edge node $s_u$ is denoted by ${D'}_u$, and the corresponding vector for all edge nodes is denoted by
$\boldsymbol{D}'$. First, the algorithm calculates the actual transmission rate of each link on the relay paths with the interference between links
considered. Then, according to the scheduling results from Algorithm \ref{algorithm2}, it obtains the data amount that can be transmitted by each link on each relay path, and the minimum value of the data amounts on the same relay path is recorded as the data amount that can be cached at the corresponding edge node.

The pseudo-code of the data updating algorithm is presented in Algorithm \ref{algorithm3}. From line 5 to line 8, the algorithm obtains the actual transmission rate of links scheduled for transmission in each time slot. From line 9 to line 15, the algorithm obtains the data amount that can be cached at each edge node. According to the scheduling results, the algorithm obtains the active time slots for each link on each relay path, and the data amount that can be transmitted by this link is calculated as in lines 13--14. The active time slots means the time slots when the link is scheduled for transmission. Then with the data amount that can be transmitted by each link on the relay path, ${D'}_{uh}$, the data amount that can be cached at the edge node $s_u$ can be obtains as ${D'_u} = \mathop {\min \{ }\limits_{{p_u^h}} {D'}_{uh}\}$, as in line 15. Finally, the caching data amount for each hotspot ${D'}_u$ is returned.

Considering lines 5--8 and
the three for loops in the algorithm, the computational complexity of the data updating algorithm is $\mathcal{O}(T|\mathbb{V}|/2+T{H_m}|\mathbb{U}|)$, which is low and can be implemented in practice.
Therefore, the overall complexity of MHRC is $\mathcal{O}({H_m}|\mathbb{U}||\mathbb{V}|^2)$ since the complexity of Algorithm \ref{algorithm2}
is much higher than that of Algorithm \ref{algorithm3}. As stated before, the probabilities of hotspots users pass by will be updated every single day, and once the scheduling decisions are made for the hotspots, caching at the edge nodes can be completed in $K$ time slots.
Thus, MHRC can be implemented offline, and update the scheduling decisions every single day, which indicates MHRC is feasible in practice.

 \SetAlFnt{\normalsize}
\begin{algorithm}\label{algorithm3}
\caption{Data Updating Algorithm}
\textbf{Input:} The set of scheduled links for each time slot $V^k$;\\
\hspace{0.95cm} The total number of time slots $K$;\\
\hspace{0.95cm} Set of relay paths $\mathbb{P}$;\\

\textbf{Initialization:}
$k=0$, $\boldsymbol{D}'=\boldsymbol{0}$;\\
\While{$k<K$}
{$k=k+1$;\\
\For{\rm{each link} $i\in V^k$}
{Obtain its transmission rate with factual interference considered, ${R'}_{i}^{k}$;}
}
\For{\rm{each node} $u \in \mathbb{U}$ }
{
   Obtain $p_u$ from $\mathbb{P}$;\\
    \For{\rm{each hop} ${p_u^h}$ on $p_u$}
    {Set ${D'}_{uh}=0$;\\
    \For{\rm{each active time slot} $k$}
     {${D'}_{uh}={D'}_{uh}+{R'}_{uh}^{k}\cdot{\Delta}$;}
     }
     ${D'_u} = \mathop {\min \{ }\limits_{{p_u^h}} {D'}_{uh}\}$;\\
}

$\mathbf{Return}$ caching data amount ${D'}_u$ for each hotspot $u$.
\end{algorithm}

\section{\label{sec:Simulation}Performance Evaluation}

In this section, we evaluate the performance of our proposed MHRC under different system settings based on real world mobility scenarios.

\subsection{Evaluation Setup}

In the evaluation, we collect 100 real world user (pedestrian) mobility trajectories in a square region of 300$m\times$300$m$ at the campus of Beijing Jiaotong University, and the probabilities of hotspots users pass by and the expected stay time of users at each hotspot are obtained from the user trajectories.
We assume the BS is located in the center of the area, and several REs are deployed in the area with Poisson distribution. In the evaluation, the widely used realistic directional antenna model is adopted, which is a main lobe of Gaussian form in linear scale and constant level of side lobes \cite{G_theta}. The gain of a
directional antenna in units of decibel (dB), denoted by $G(\theta )$, can be calculated as,
\begin{equation}
G(\theta ) = \left\{ {\begin{array}{*{20}{c}}
{{G_0} - 3.01 \cdot {{(\frac{{2\theta }}{{{\theta _{ - 3{\rm{dB}}}}}})}^2},\;0^ \circ \le \theta  \le {\theta _{ml}}/2};\\
{{G_{sl}},\hspace{2.2cm}{\theta _{ml}}/2 \le \theta  \le {{180}^ \circ }},
\end{array}} \right.
\end{equation}
where $\theta$ denotes an arbitrary angle within the range $[0^ \circ, 180 ^ \circ]$, ${{\theta _{ - 3{\rm{dB}}}}}$ denotes the angle of the half-power beamwidth, and ${{\theta _{ml}}}$ denotes the main lobe width in units of degrees. The relationship between ${{\theta _{ml}}}$ and ${{\theta _{ - 3{\rm{dB}}}}}$ is ${\theta _{ml}} = 2.6 \cdot {\theta _{ - 3{\rm{dB}}}}$.
${{G_0}}$ is the maximum antenna gain, and can be obtained by ${G_0} = 10\log {(\frac{{1.6162}}{{\sin ({\theta _{ - 3{\rm{dB}}}}/2)}})^2}$. The side lobe gain, ${{G_{sl}}}$, can be expressed as ${G_{sl}} =  - 0.4111 \cdot \ln ({\theta _{ - 3{\rm{dB}}}}) - {\rm{10}}{\rm{.579}}$. The default evaluation parameters are summarized in Table \ref{tab:para-MHRC}, and we investigate the performance of MHRC under different system parameters.


\begin{table}[htbp]
\begin{center}
\caption{Simulation Parameters}
\def \temptablewidth {0.9\textwidth}
\begin{tabular}{ccc}
\hline
\textbf{Parameter}&\textbf{Symbol}&\textbf{Value}\\
\hline
System bandwidth & W & 2160 MHz \\
Background noise &$N_0$& -134dBm/MHz\\
Path loss exponent & $\gamma $ & 2\\
Transmission power & $P_{t}$ & 30 dBm\\
Time slot duration &$\Delta $& 1 s\\
Half-power beamwidth & ${{\theta _{ - 3{\rm{dB}}}}}$ & ${\rm{30}}^\circ $\\
Efficiency of the transceiver design &$\eta$ & 0.5\\
Shrinking parameter & $\beta$ & $0.9$ \\
Maximum number of hops & $H_m$ & 8\\
Interference threshold & $\sigma$ & $10^{-10}$\\
Rely node number &$|\mathbb{V}|$ & 30\\
Number of time slots & $K$ & 5400\\
\hline
\end{tabular}
\label{tab:para-MHRC}
\end{center}
\end{table}

In this evaluation, we compare our proposed MHRC with other two schemes to show the advantages of multi-hop D2D caching and concurrent transmissions in our scheme.

1) \emph{\textbf{Unicast}}: In the unicast scheme, the BS transmits to the user directly when the user passes by and stays at the hotspots.

2) \emph{\textbf{CachUni}}: In the CachUni scheme, caching is adopted, but the cached data is transmitted directly from the BS to the edge node.
The transmissions of caching for the hotspots with the higher probability are scheduled first.

Considering the objective function of (\ref{obj2}), the performance metric in our evaluation is the expected amount of data cached during the users passing by, i.e., $E(\sum\limits_{u \in {U_0}} {{D_u}} )$. In the results below, this metric is denoted by E(D) in the unit of bit. Besides, we also evaluate the energy efficiency of different schemes, which is the expected amount of data cached divided by the total consumed energy consumption.

\subsection{Result Analysis}


\begin{figure}[tbp]
\begin{minipage}[t]{1\linewidth}
\centering
\includegraphics[width=0.85\columnwidth]{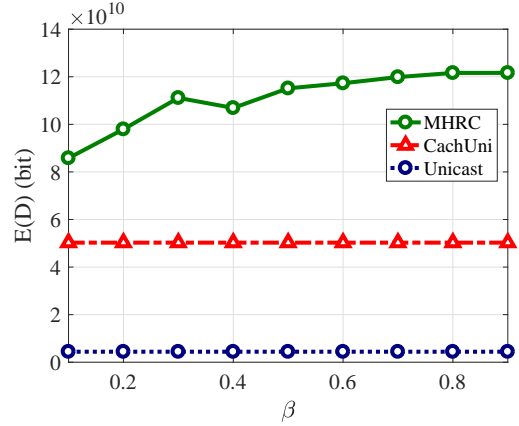}
\end{minipage}%
\caption{The expected cached data of the three schemes under different shrinking factor $\beta$.}
\label{fig:D-beta} 
\vspace*{-3mm}
\end{figure}

In Fig. \ref{fig:D-beta}, we plot the expected cached data of the three schemes under different $\beta$. From the results, we can observe that MHRC achieves the best performance in terms of the expected caching data amount. The gap between MHRC and the CachUni scheme is due to the multi-hop D2D caching and concurrent transmission mechanisms in our scheme. For the unicast scheme, where caching is not enabled, the performance is even worse. Since the shrinking factor $\beta$ represents the ratio of remaining data amount after the original data amount cannot be transmitted successfully in Algorithm \ref{algorithm2},
when $\beta$ is small, the estimated data amount for each edge node decreases too quickly, and the achieved caching data amount is low. Thus, with the increase of $\beta$, the expected caching data amount of MHRC increases. When $\beta$ increases to 0.7, the system approaches to saturation, and the performance keeps stable since the data amount will always decrease to the saturation value.
When $\beta$ is equal to 0.9, our scheme improves the
expected cached data amount by about 142\% compared with the CachUni scheme.

\begin{figure}[tbp]
\begin{minipage}[t]{1\linewidth}
\centering
\includegraphics[width=0.85\columnwidth]{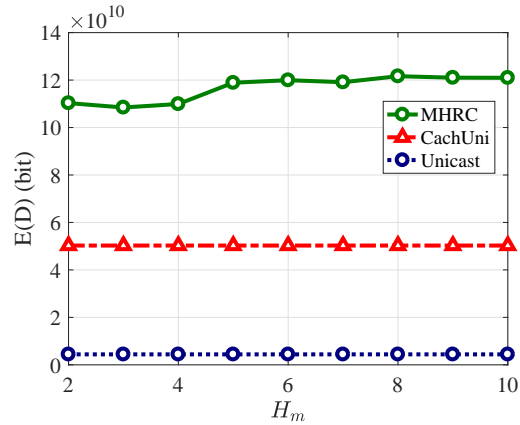}
\end{minipage}%
\caption{The expected cached data of the three schemes under different maximum hop numbers.}
\label{fig:D-Hm} 
\vspace*{-3mm}
\end{figure}

In Fig. \ref{fig:D-Hm}, we plot the expected cached data of the three schemes under different maximum hop numbers.
We can also observe that our scheme achieves the best performance among three schemes. With the increase of the maximum hop number,
our scheme achieves higher expected caching data amount, which indicates multi-hop D2D relaying improves the performance. When the maximum
hop number increases to 5, the system performance approaches the saturation since the number of time slots is limited and more hops means more links to be scheduled.

\begin{figure}[tbp]
\begin{minipage}[t]{1\linewidth}
\centering
\includegraphics[width=0.85\columnwidth]{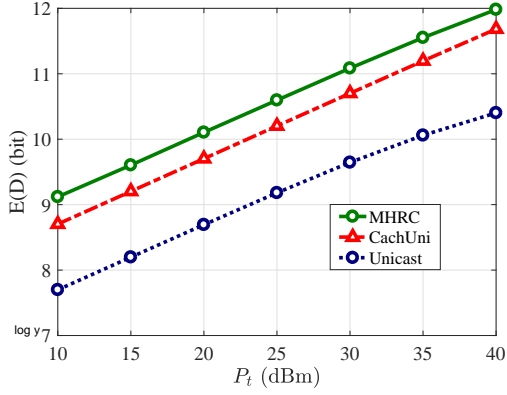}
\end{minipage}%
\caption{The expected cached data of the three schemes under different transmission power.}
\label{fig:D-Pt} 
\vspace*{-3mm}
\end{figure}

In Fig. \ref{fig:D-Pt}, we plot the expected cached data of the three schemes under different transmission power.
To show the performance gap between different schemes more clearly, we show the results with the Y-axis using the logarithmic coordinates.
We can also observe the superior performance of MHRC compared with other two schemes.
The expected caching data amount increases with the transmission power, which is obvious since more transmission power leads to higher transmission ability, and more data can be cached at the edge node.

\begin{figure}[tbp]
\begin{minipage}[t]{1\linewidth}
\centering
\includegraphics[width=0.85\columnwidth]{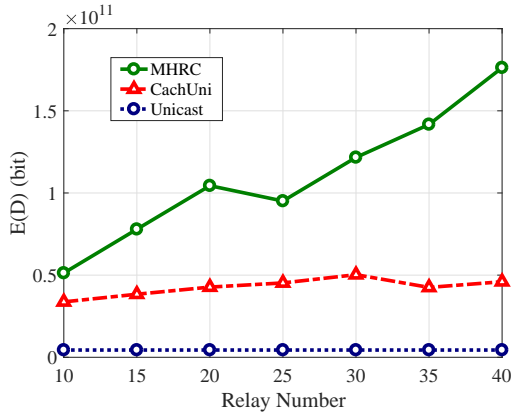}
\end{minipage}%
\caption{The expected cached data of the three schemes under different relay numbers.}
\label{fig:D-RN} 
\vspace*{-3mm}
\end{figure}

In Fig. \ref{fig:D-RN}, we plot the expected cached data of the three schemes under different relay numbers.
Generally, we can observe that MHRC achieves higher expected caching data amount with more relay nodes deployed. With more relay nodes deployed, the distance between relay nodes is shorter on average, and higher transmission rates of links can be achieved.
Since the relay number only has impact on the location of the edge nodes in CachUni, the performance of CachUni is not affected by the relay number too much.
When the number of relay nodes is 40, our scheme improves the expected cached data amount by about 283\% compared with the CachUni scheme.

\begin{figure}[tbp]
\begin{minipage}[t]{1\linewidth}
\centering
\includegraphics[width=0.85\columnwidth]{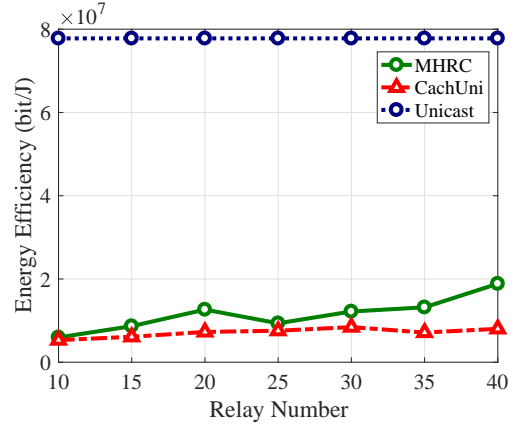}
\end{minipage}%
\caption{The energy efficiency of the three schemes under different relay numbers.}
\label{fig:E-RN} 
\vspace*{-3mm}
\end{figure}

In Fig. \ref{fig:E-RN}, we plot the energy efficiency of the three schemes under different relay numbers.
From the results, we can observe that the unicast scheme achieves the highest energy efficiency.
This is because without multihop relaying transmission and caching, the unicast scheme consumes less energy
although less cached data amount is achieved. Compared with the CachUni scheme, our scheme can achieve higher
energy efficiency, and the results demonstrate the superior performance of our scheme among caching based schemes.


\begin{figure}[tbp]
\begin{minipage}[t]{1\linewidth}
\centering
\includegraphics[width=0.85\columnwidth]{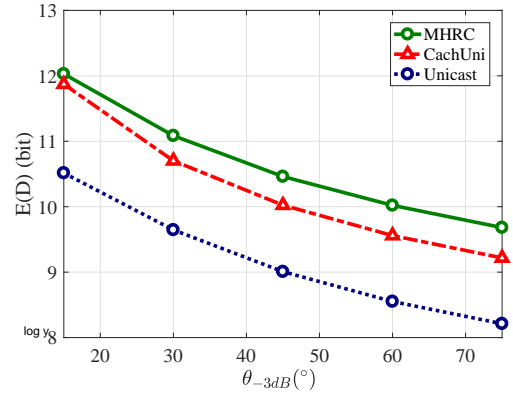}
\end{minipage}%
\caption{The expected cached data of the three schemes under different ${{\theta _{ - 3{\rm{dB}}}}}$.}
\label{fig:D-theta} 
\vspace*{-3mm}
\end{figure}

In Fig. \ref{fig:D-theta}, we plot the expected cached data of the three schemes under different ${{\theta _{ - 3{\rm{dB}}}}}$.
We can observe that our proposed MHRC achieves the best performance in terms of the expected caching data amount among three schemes.
When ${{\theta _{ - 3{\rm{dB}}}}}$ increases, the antenna gain decreases, and the transmission rate decreases. Consequently, the expected data amount
decreases.

\begin{figure}[tbp]
\begin{minipage}[t]{1\linewidth}
\centering
\includegraphics[width=0.85\columnwidth]{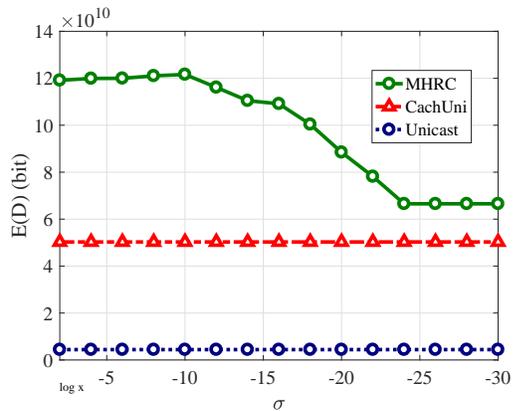}
\end{minipage}%
\caption{The expected cached data of the three schemes under different thresholds $\sigma$.}
\label{fig:D-sigma} 
\vspace*{-3mm}
\end{figure}

We plot the expected cached data of the three schemes under different interference thresholds $\sigma$ in Fig. \ref{fig:D-sigma}.
The interference threshold $\sigma$ is shown with the logarithmic coordinates.
Since concurrent transmissions are disabled in the CachUni and the Unicast scheme, their performance does not change with $\sigma$.
For MHRC, when the interference threshold is small, the conditions for concurrent transmissions are rigid, and the advantages of concurrent transmissions are not exploited fully. Thus, the expected caching data amount is relatively low. With higher interference threshold,
more concurrent transmissions can be enabled to improve performance, which leads to the increase of the expected caching data amount in the curve of
MHRC. When the interference threshold increases further, the advantages brought by concurrent transmissions are counteracted by the interference between links, and the performance of MHRC becomes stable. To achieve a relatively high performance, the default value of $\sigma$ is $10^{-10}$
as in Table \ref{tab:para-MHRC}.

\begin{figure}[tbp]
\begin{minipage}[t]{1\linewidth}
\centering
\includegraphics[width=0.85\columnwidth]{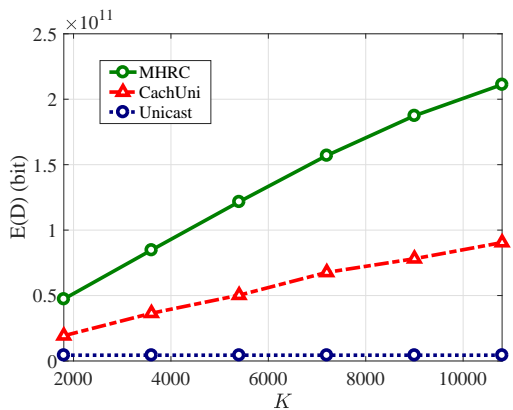}
\end{minipage}%
\caption{The expected cached data of the three schemes under different number of time slots $K$.}
\label{fig:D-K} 
\vspace*{-3mm}
\end{figure}

In Fig. \ref{fig:D-K}, we plot the expected cached data of the three schemes under different number of time slots $K$.
We can observe that our scheme achieves the best performance, and the expected caching data amount increases with the number of time slots $K$.
With more time slots available, more transmissions for caching at the edge nodes can be scheduled, and higher caching data amount can be transmitted.
When the number of time slots is 10800, our scheme improves the expected cached data amount by about 133\% compared with the CachUni scheme.

\begin{figure}[tbp]
\begin{minipage}[t]{1\linewidth}
\centering
\includegraphics[width=0.85\columnwidth]{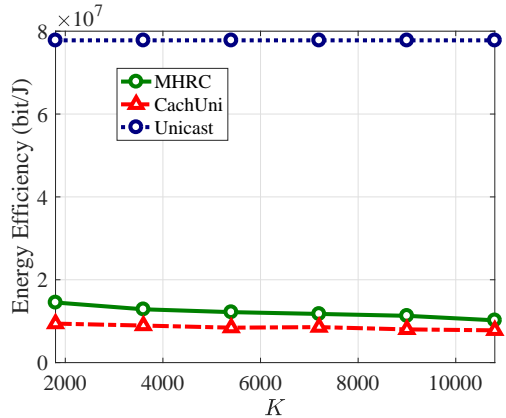}
\end{minipage}%
\caption{The energy efficiency of the three schemes under different number of time slots $K$.}
\label{fig:E-K} 
\vspace*{-3mm}
\end{figure}

In Fig. \ref{fig:E-K}, we plot the energy efficiency comparison of the three schemes under different number of time slots $K$.
We can observe that the results are consistent with those in Fig. \ref{fig:E-RN}. Compared with the CachUni scheme, our scheme also achieves
better energy efficiency.

\subsection{Error Analysis}

\begin{figure}[tbp]
\begin{minipage}[t]{1\linewidth}
\centering
\includegraphics[width=0.85\columnwidth]{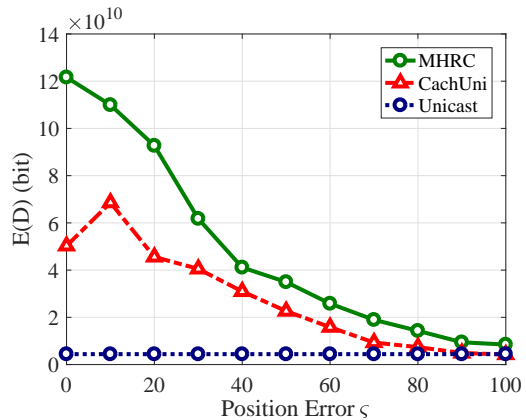}
\end{minipage}%
\caption{The expected cached data of the three schemes under different location error.}
\label{fig:D-E} 
\vspace*{-3mm}
\end{figure}

Since our scheme relies on the trajectories of users, locations of the hotspots will have impact on the performance.
In this part, we investigate the impact of location error of the hotspots on the performance.
We assume the location error $(\Delta x,\Delta y)$ is added to the locations of hotspots,
and $(\Delta x,\Delta y)$ is subject to two-dimensional normal distribution $N\left( {\left( {\begin{array}{*{3}{c}}
0\\
0
\end{array}} \right),\left( {\begin{array}{*{3}{c}}
{\begin{array}{*{3}{c}}
\varsigma &0
\end{array}}\\
{\begin{array}{*{3}{c}}
0&\varsigma
\end{array}}
\end{array}} \right)} \right)$, where $\Delta x$ is the position error in X-axis, and $\Delta y$
is the position error in Y-axis, respectively. In Fig. \ref{fig:D-E}, we plot the expected cached data of the three schemes under different location error. We can observe that the expected cached data decreases with the position error globally, which is consistent with intuitions.
Besides, our scheme outperforms the CachUni scheme when $\varsigma$ is less than 100. Since the unicast scheme transmits to the user directly
without using the positions of hotspots, its performance doesn't change with the position error. Thus, our scheme
is robust to the position error.

\begin{figure}[tbp]
\begin{minipage}[t]{1\linewidth}
\centering
\includegraphics[width=0.85\columnwidth]{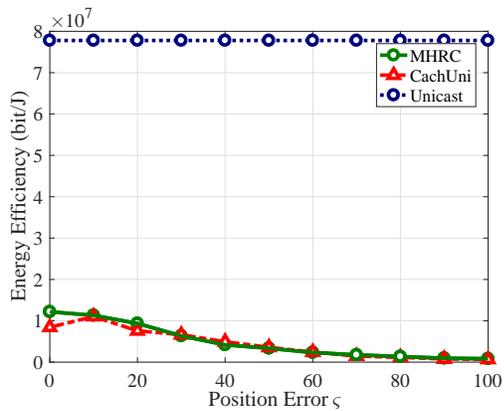}
\end{minipage}%
\caption{The energy efficiency of the three schemes under different location error.}
\label{fig:E-E} 
\vspace*{-3mm}
\end{figure}

In Fig. \ref{fig:E-E}, we plot the energy efficiency of the three schemes under different location error.
We can observe that the energy efficiency decreases with the position error globally, which is also consistent with intuitions.
When the position error is less than 100, our scheme always performs better than the CachUni scheme.

In summary, the above extensive results obtained in the performance evaluation demonstrate our scheme MHRC is able to achieve significantly higher
expected caching data amount compared with other two schemes. In terms of energy efficiency, our scheme still outperforms the CachUni scheme.
The key mechanisms of our scheme are the utilization of multi-hop D2D relaying
and concurrent transmissions for caching with the user mobility trajectories considered.


\section{\label{sec:Conclusion}Conclusion}

In this paper, we proposed a mobility-aware and multihop-D2D-relaying-based scheduling scheme for caching in mmWave band of fog computing systems, called MHRC, where
multi-hop D2D paths are established for edge nodes near the hotspots, and concurrent transmissions are exploited to improve the system performance.
We conduct extensive performance evaluations under various system parameters, and the results demonstrate MHRC achieves
more than 1x higher expected cached data amount compared with other existing schemes. Due to weak diffraction ability, mmWave links suffer the blockage problem from obstacles, and we will address the blockage problem in the fog computing system in the future.


\end{document}